\documentclass{epl}
\usepackage[dvips]{epsfig}
\usepackage{subfigure}
\usepackage{graphicx}
\title{Non-linear terms in 2D cosmology}
\author{F. P.
Devecchi\footnote{devecchi@fisica.ufpr.br}\and
M. L. Froehlich}
\institute{Departamento de F\'\i sica, Universidade Federal do Paran\'a,
Caixa Postal 19044, 81531-990, Curitiba, Brazil[may 2005]}

\def\be{\begin{equation}}
\def\ee#1{\label{#1}\end{equation}}
\begin{document}
\maketitle

\begin {abstract}
  
In this work we investigate the behavior of two-dimensional (2D)
 cosmological
models, starting with the  Jackiw-Teitelboim (JT) theory of gravitation.  A geometrical term, 
non-linear in the scalar curvature $R$,
is added to the JT dynamics to test if it could play the role of dark energy in a
2D  expanding universe. This formulation makes possible, first, the description
 of an early (inflationary) 2D universe, when 
  the van der Waals (vdW) equation of state  is used to construct the 
energy-momentum tensor of the gravitational sources.
 Second, it is found that for 
later times the 
non-linear term  in $R$ can generate 
an old 2D universe in accelerated expansion, where  
an ordinary matter dominated 
era  evolves into a decelerated/accelerated transition, giving to the dark energy 
effects a geometrical origin. The results emerge through numerical analysis, following  the evolution in time of 
the scale
factor, its acceleration, and the energy densities of constituents. 

\end{abstract}

\section{Introduction}

The study of theories of gravity in lower dimensions has 
received extensive consideration in the literature \cite{Brown}. 
Taking the particular case of two-dimensional (2D) theories,  the gravity
 formulation proposed by   
  Teitelboim  and Jackiw \cite{Teit} (JT model) provided 
consistent results at the  classical and quantum levels \cite{Brown}. 
This model   is the most imediate substitute   of general
relativity in 2D field formulations, since the Einstein theory furnish no 
dynamics in 2D \cite{Brown}. 
The JT model was under investigation also in a cosmological context
\cite{Cornish, Deve}. 
 The results obtained include 
the description of a Universe 
filled with ordinary  matter or/and electromagnetic
radiation, an inflationary 2D universe and a three-eras universe where a 
cosmological constant was responsible for a final accelerated period, that could de associated
to the effects of dark energy
\cite{Cornish}. 

In fact, the dark energy problem, as a fundamental discussion in cosmology, has been analised using different 
approaches. Recently, it was proposed   a geometrical 
formulation for the dark energy behavior in a 4D universe\cite{Vol}. This
idea comes in substitution of the standard formulations that takes the dark energy
as an usual source\cite{Vol}. This geometrical formulation has
as central point a modification in the Einstein action that includes a 
non-linear term in the scalar curvature $R$, that would be key ingredient behind the present accelerated 
expansion in the universe\cite{Vol}.

The proposal of our  work is to make an analogous modification in the JT model,
and analyse  in which cases     the non-linear term in $R$ is responsible 
for a final period of positive acceleration. In those cases the geometrical 
term would be a substitute to the cosmological constant  in playing the role 
of dark energy in 2D universes. Following our  previous effort \cite{Deve},
we model the inflaton-matter source using the van der Waals equation of state
(vdW), that is responsible for an initial strongly accelerated period  (first period) corresponding to an 
inflationary young 
2D universe (dominated by the inflaton). As the system evolves
the vdW equation approaches a barotropic equation of state (modelling  a period where
matter/radiation  start to predominate over the inflaton contribution), producing  a decelerated expansion
of the 2D universe (second period). In these two periods  the non-linear  term in $R$ gives no sensible 
contribution to the dynamics, and the 
standard
JT results\cite{Deve} are reproduced. It is in a final (third) period  that the non-linear term
 promotes, in some situations, an accelerated  era  as a geometrical effect; so  the usual negative-pressure
sources are not needed to simulate the dark energy dominated period. In the following section we make a 
brief review
of the JT model in cosmology that works as preface to the  presentation of our results.

\section{The Jackiw-Teitelboim model and cosmology}
\indent
As it was mentioned in the introduction  the Jackiw-Teitelboim 
(JT) model\cite{Teit}   
is a succesful formulation of 2D gravity. 
Among other results, that include a consistent quantum version, the model is able to furnish interesting
results in 2D cosmologies 
\cite{Cornish, Deve}. The dynamics of the JT theory 
is ruled by
\be
R (x) = -\kappa T^\mu_\mu(x)+\Lambda \, ,
\ee{RW21a}
\noindent where $R$ is curvature scalar, $T^{\mu \nu}$ is the energy 
momentum tensor of 
the sources, and $\Lambda$ is a cosmological constant. 
The proposal of the  JT model (in substitution to  the Einstein field equations that in 2D are obsolete\cite{Brown})  
focused on the fact that  the full geometrical 
information for the 2D space-time is encapsulated in the curvature scalar $R$ and as a consequence 
the gravitational field dynamics
can be put in a scalar form. 
The above mentioned cosmological applications 
consider an  hypothetical 2D Universe where the 
assumptions of spatial homogeneity 
and isotropy are  invoked. The
Robertson-Walker metric has  the
following form in a 2D Riemannian space-time 
\be 
ds^2=(dt)^2-a(t)^2(dx)^2,
\ee{RW1a}
where  
$a(t)$  is the  cosmic scale factor, that encapsulates
the complete information
about the  evolution of the gravitational field created by the sources. 
The equations of motion (together with the
conservation law ${T^{\mu\nu}}_{;\nu} =0$)
 can be expressed as (making $\kappa=2\pi G_1$ with $G_1=1$)
\be
{\ddot a\over a}=-4\pi(\rho-p)+{\Lambda\over2}\, ,\,\,\,\,\,\,\,\,
\dot \rho+{\dot a\over a}(\rho+p)=0 \, ,
\ee{99}
\noindent where $\rho$ and $p$ are  the energy 
density and the hidrostatic pressure of the sources, respectively.  
The cosmological solutions of the JT model  include an old universe filled with 
radiation and/or
matter~\cite{Cornish} (a 2D counterpart of the results found in \cite{Mur}); 
inflationary scenarios (with transitions to a matter era) are also possible, when the cosmic fluid obeys the 
van der
Waals (vdW) equation of state\cite{Deve}. Another interesting result is that a three eras regime (a first 
strongly accelerated period followed by a decelerated one dominated by matter/radiation; and a final accelerated 
era)
 can be obtained 
when a cosmological constant $\Lambda$ is included. Here this constant is playing the  role 
of a dark energy density  that must always have negative values  ( a situation that occurs only in 2D \cite{Deve}).

\section{The $1/R$ correction in 2D}

Our proposal is that 
The JT model dynamics   
be corrected by a term 
of the form   $\omega ^2 /R$. This dynamics is inspired by the recent 
investigations 
\cite{Vol}  of giving a {\it geometrical}  origin to the   dark energy
constituent  in 4D cosmological models. Here
the contribution of the non-linear term becomes important for later times, 
promoting an acelerated expansion period, making
possible, at least qualitatively, the association with the dark energy 
contributions. In the 2D case we want to investigate the effects produced 
by the 
non-linear
term in $R$ and in which cases this can be associated to a dark-energy
contribution. The field equation read in this case 

\begin{equation}
R - \frac{\omega ^2}{R} + \kappa T = 0\,  \, \,
\Longrightarrow \, \, \,
R=\frac{-\kappa T \pm \sqrt{\kappa ^2 T^2 +4\omega ^2}}{2}\,\, ,
\label{R}
\end{equation}

\noindent where the non-linear term is controlled by 
an $\omega$ parameter. The sources are supposed to 
be ruled by a perfect fluid energy-momentum tensor
$T^{\mu \nu} = p g^{\mu \nu} + (p + \rho)U^{\mu}U^{\nu}
$. The choose of the sign in the field equation (4) is related to the 
preservation of the JT limit\cite{Deve}; to guarantee the existence of this limit one must follow
 the sign of the energy-momentum trace $T$ during the evolution of the system. 
The scalar curvature  for the metric (2) in given by 
$R=\frac{2\ddot a}{a}$\cite{Deve}.
Using these expressions in the field equation (4) (together with 
the energy-momentum tensor
conservation law) 
we obtain the following system of equations
\begin{equation}
\frac{2\ddot{a}}{a}=\frac{-\kappa (\rho-p) \pm \sqrt{\kappa ^2 (\rho-p)^2 +
4\omega ^2}}{2}, \quad \quad  \dot{a}(\rho+p)+a\dot{\rho}=0\, ,
\label{acel2d}
\end{equation}

\noindent where the gravitational source ( a mixture of inflatons and matter) 
is 
supposed to obey \cite{Deve} the van 
der Waals (vdW) equation 
of state 
\begin{equation}
p=\frac{b\rho}{1-\alpha \rho}\, ,
\label{vdw}
\end{equation}
where $b$  and $\alpha $ are constant parameters; the use of the vdW equation
in 
cosmological 
models was proposed in \cite{Capo1} and used in 2D models
in \cite{Deve}.

The differential equations system (5) can be solved numerically. As a 
starting point, the 
following (normalized) initial conditions were used
 $a(0)=1$, $\dot{a}(0)=1$ e $\rho(0)=1$;  these conditions simulate a 2D 
young Universe, at the beginning of an strongly accelerated
(inflationary) period\cite{Deve}. As we mentioned before the choice of sign in 
equations (4) depends on the behavior of  $T$. What we verify is that the sign of $T$ separate naturally  the 
time interval in two  periods (being negative in the first period and positive in the second).

\begin{figure}[h]
  \centering
  \mbox{
    \subfigure[]
    {\epsfig{figure=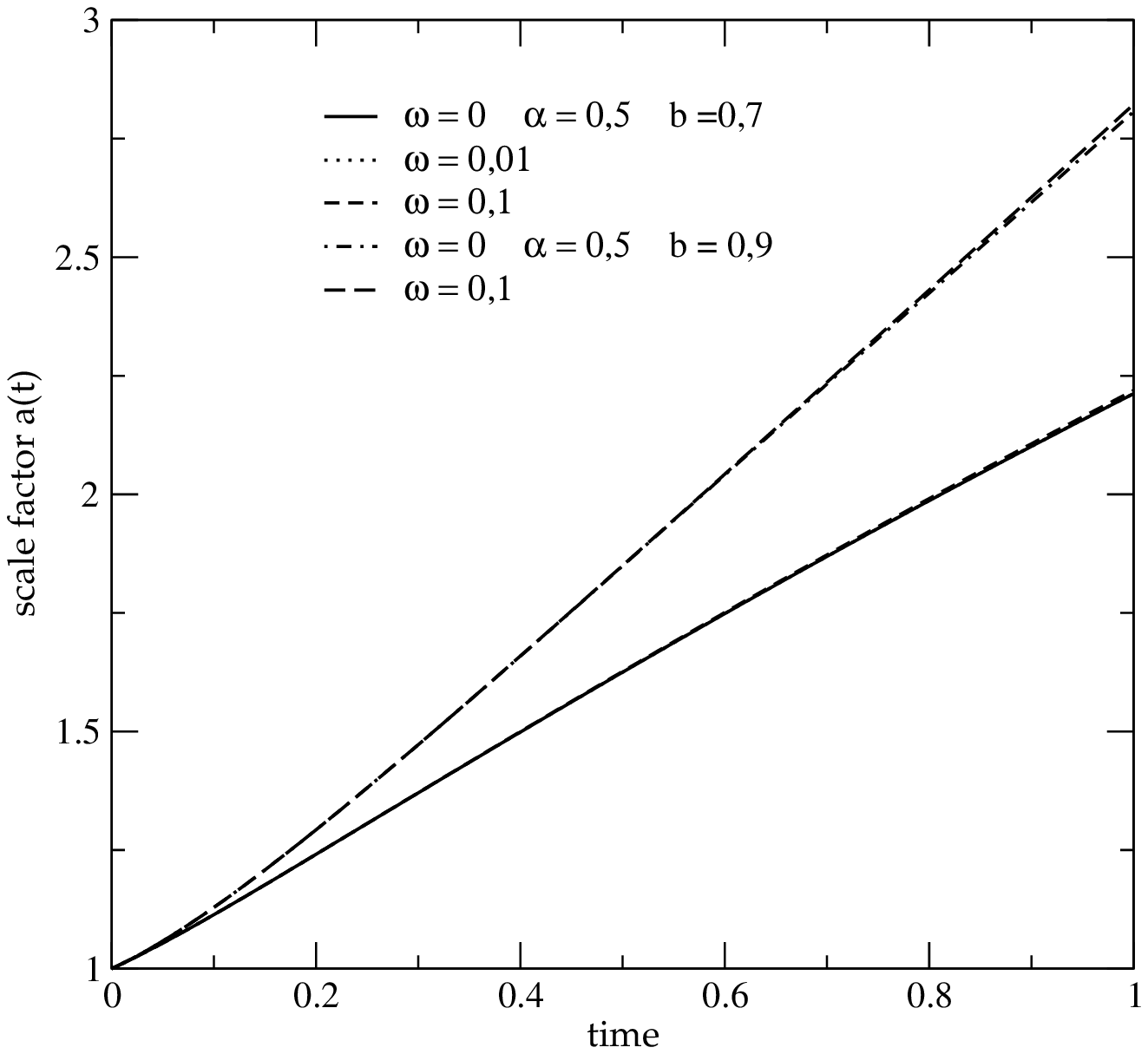,height=5.6cm,width=6.6cm}}
    \subfigure[]
    {\epsfig{figure=101.eps,height=5.3cm,width=6.5cm}}
    }\\
 \caption{Evolution in time for scale factor a(t), for different values of
the parameters, first period.}
\label{fig1}
\end{figure}

In figure (\ref{fig1})  we present the evolution 
in time of the scale 
factor $a(t)$ for
different values of the parameters.  The first  result observed    is     
that the expansion 
becomes faster when either the parameters $b$  
or $\alpha$ are increased, which is in accord with the behavior verified in
\cite{Deve}. More importantly,  
the results show that  the presence of the 
non-linear term (using different values of the control parameter $\omega$) 
does not 
affect the expansion (notice the superimposition of 
curves in figures (a) and (b); the term in $1/R$ is in fact buried by the 
presence of the energy-momentum trace term. This is in accord with the idea 
that the non-linear term play the role of the 2D dark energy constituent, 
becoming
important only at later times (second period). 

\begin{figure}[h]\vskip0.3truecm\begin{center}
\includegraphics[height=5.5cm, width=7.0cm]{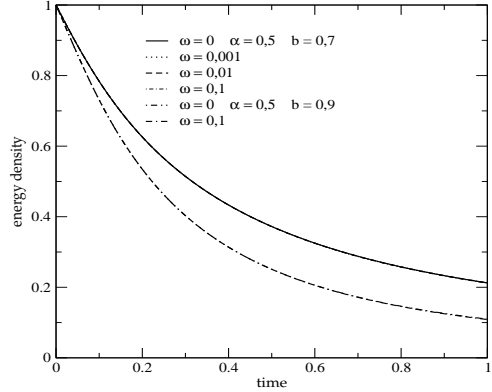}
\caption{Evolution in time for the vdW energy density, 
taking different values of
the parameters, first period.}
\label{fig2}
\end{center}
\end{figure}

In figure (\ref{fig2}) we display the evolution of the energy density 
$\rho(t)$ of the vdW 
constituent. As expected, 
considering the behavior of the scale factor, the function $\rho(t)$ is in a 
ever 
decreasing regime; increasing values of $b$ (and $\alpha$, not displayed here)
 promote a faster 
fall on the values of $\rho(t)$. Changes in $\omega$ don't 
modify the behavior of the energy density for the 
early 2D  universe. 

\begin{figure}[h]
  \centering
  \mbox{
    \subfigure[]
    {\epsfig{figure=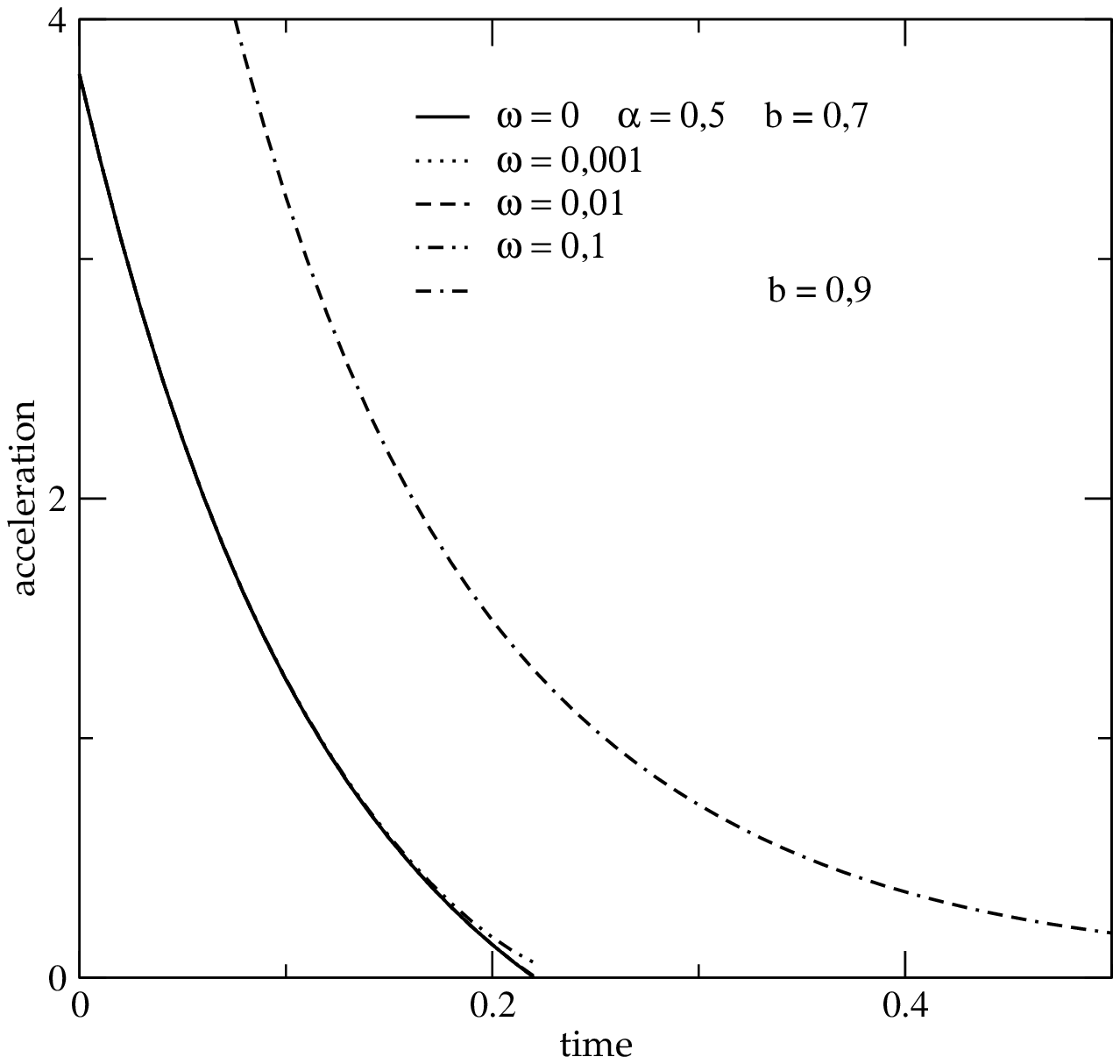,height=5.0cm,width=6.6cm}}
    \subfigure[]
    {\epsfig{figure=4.eps,height=4.8cm,width=6.6cm}}
    }\\
 \caption{Evolution in time for acceleration $\ddot{a}(t)$ 
taking different values of
the parameters, first period.}
\label{fig3}
\end{figure}

The acceleration $\ddot a(t)$, as shown in figure (\ref{fig3}) (a) and (b), 
is also not 
modified by the 
presence of the non-linear term in $R$. Increasing values of $\alpha $ 
turn the fall
of the curve more drastic (together with an increasing value of the initial 
acceleration $\ddot a(0)$), approaching the turning point to a decelerated 
regime where matter starts to predominate (as  the vdW equation of states 
approaches a barotropic equation as time increases).

\begin{figure}[h]
  \centering
  \mbox{
    \subfigure
    {\epsfig{figure=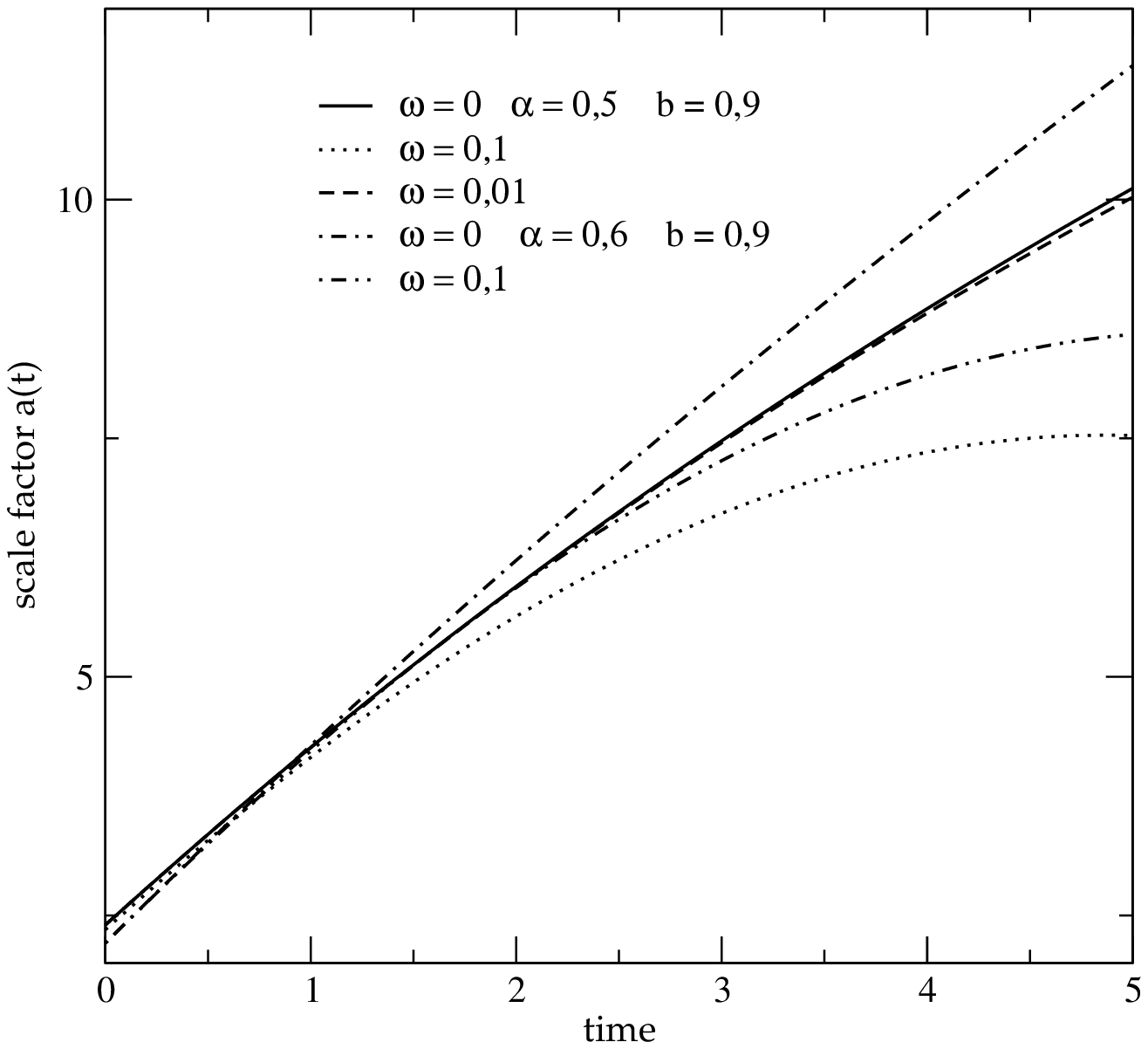,height=4.9cm,width=6.9cm}}
    \subfigure
    {\epsfig{figure=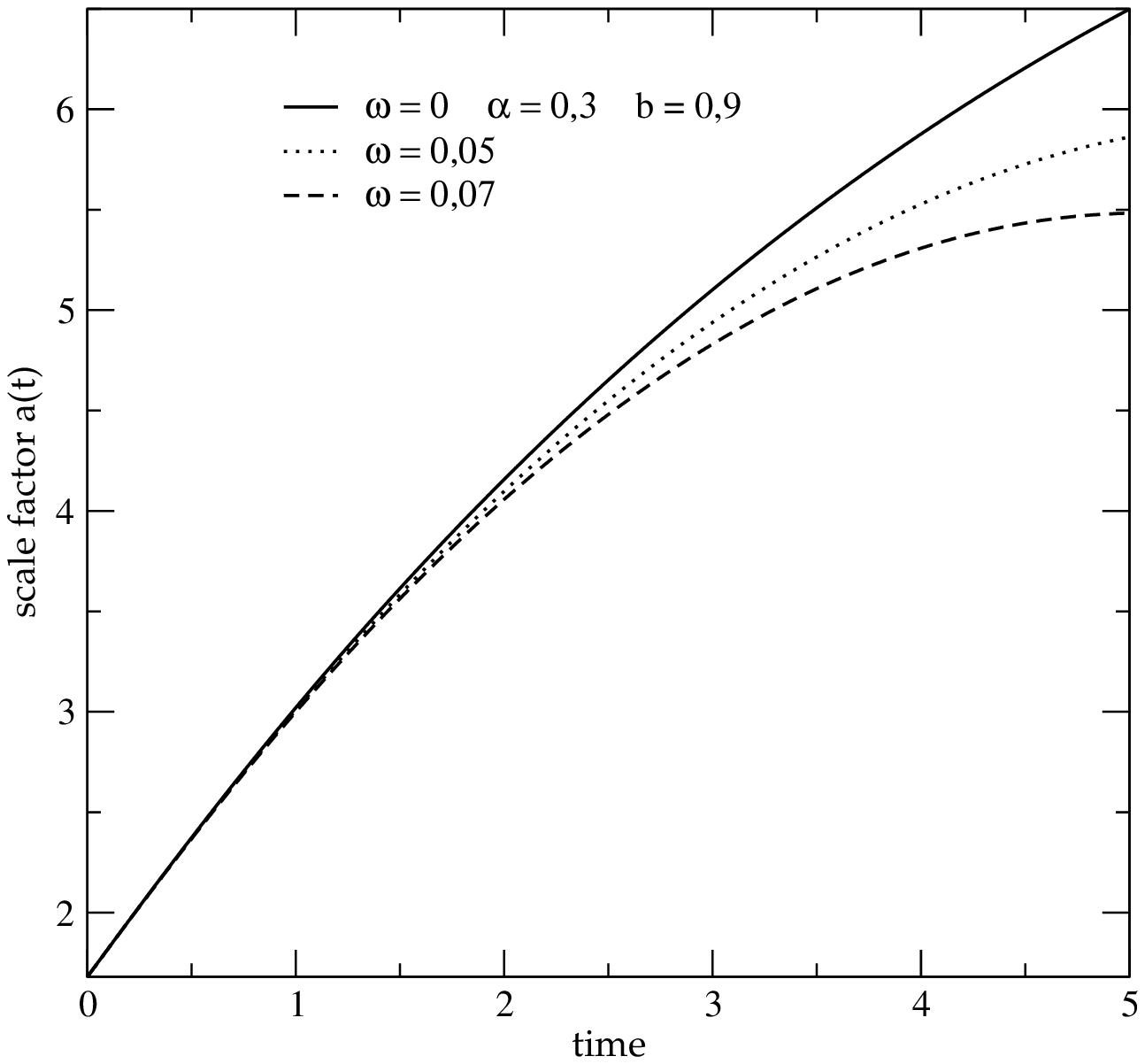,height=4.9cm,width=6.9cm}}
    }\\
 \caption{Evolution in time for scale factor a(t) taking different values of
the parameters, second period.}
\label{fig4}
  \centering
  \mbox{
    \subfigure[]
    {\epsfig{figure=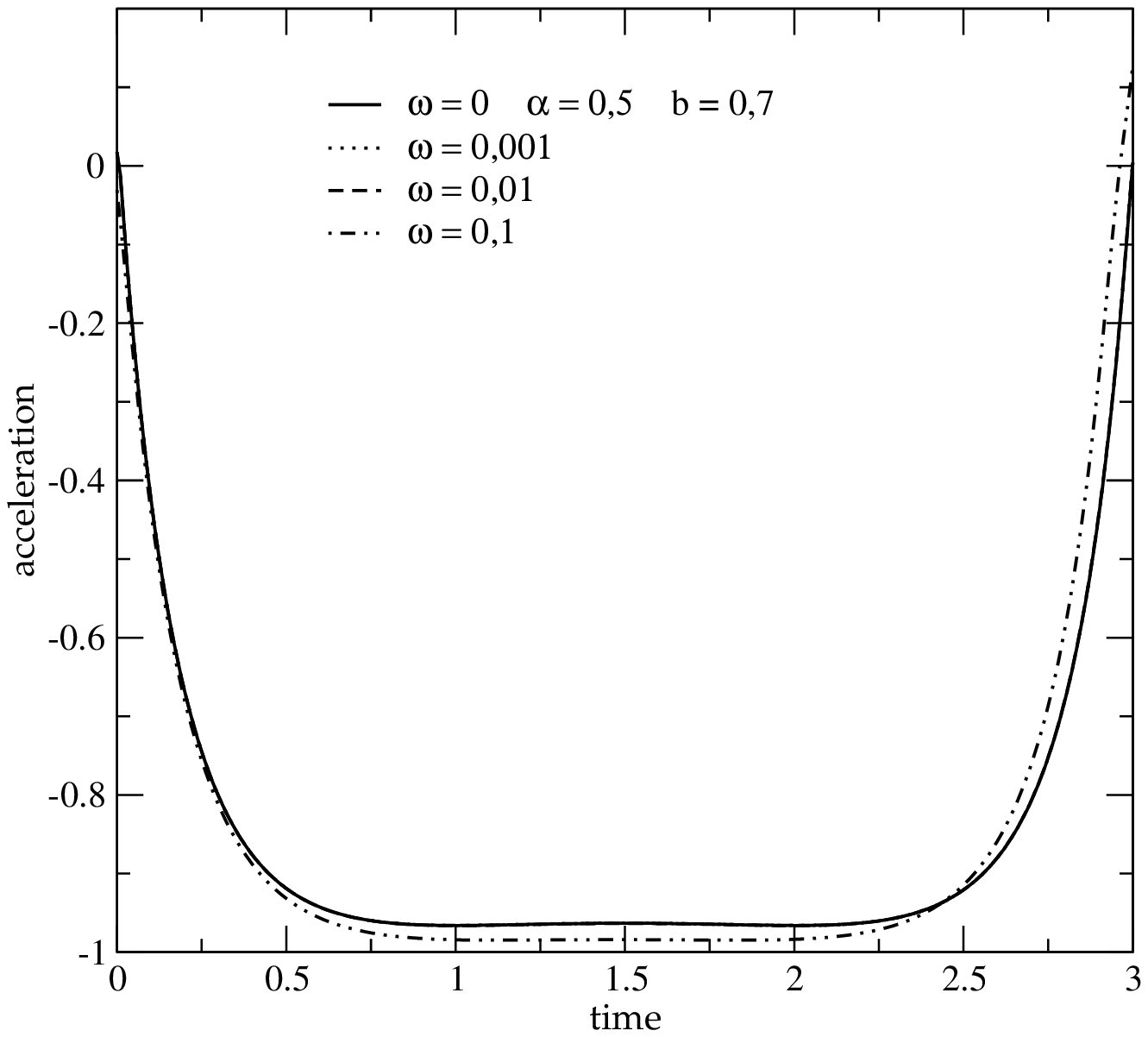,height=5.5cm,width=6.9cm}}
    \subfigure[]
    {\epsfig{figure=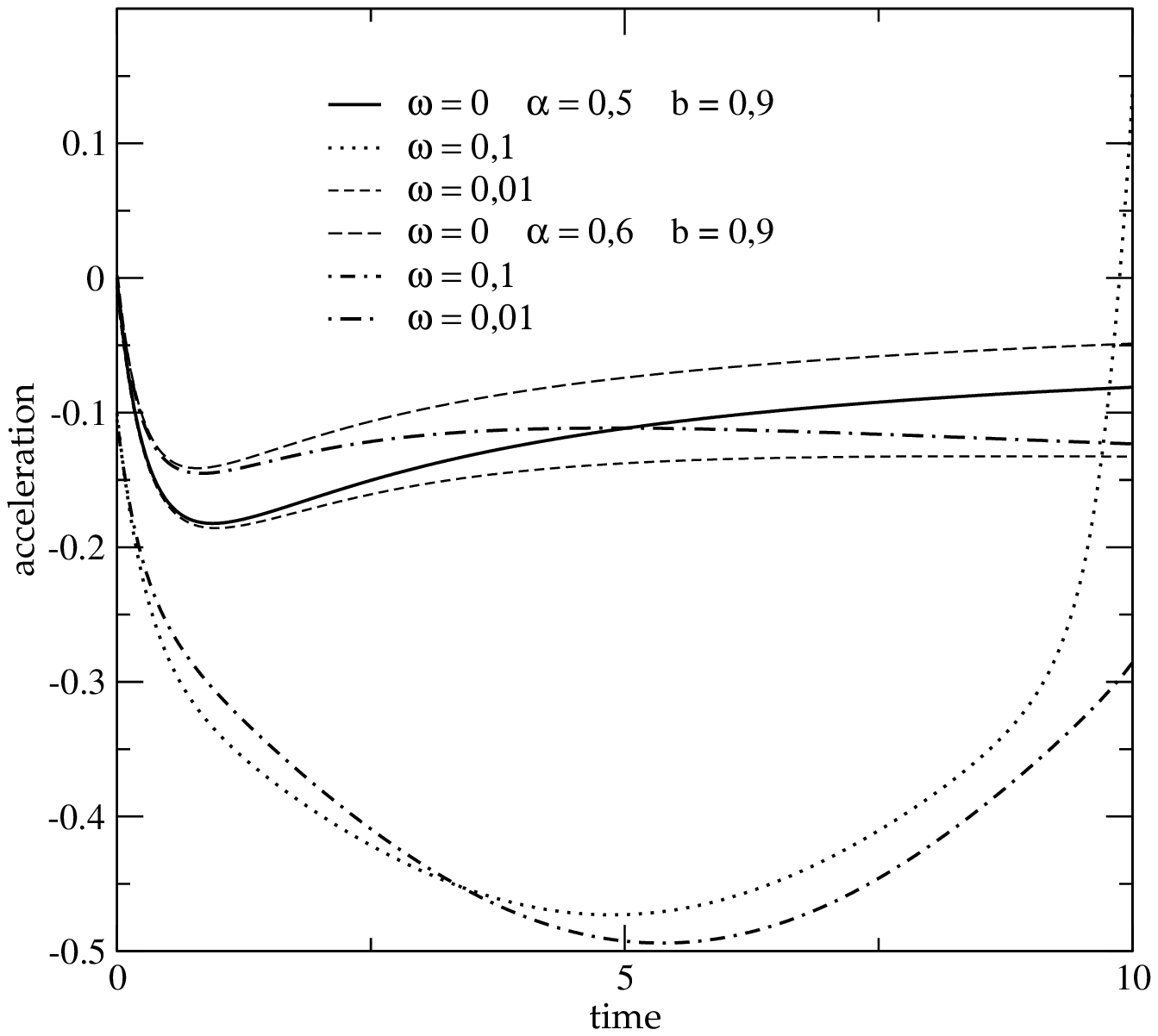,height=5.5cm,width=6.9cm}}
    }\\
 \caption{Evolution in time for acceleration $\ddot{a}(t)$ taking 
different values of
the parameters, second period.}
\label{fig5}
\end{figure}

In figures (\ref{fig4}) and (\ref{fig5})  we display 
the behavior of the scale factor $a(t)$,
the vdW energy density and the acceleration $\ddot a(t)$ for the second period
(when $T>0$). The curves 
show  an ever expanding 2D universe (in some cases  a {\it big crunch } 
situation emerges 
but only for  later times ($t\approx 10$)), with increasing values of 
$\alpha$ and 
$b$ furnishing a faster expansion, as it was in the case of the first period. 
 An expected behavior 
for the vdW energy density is also obtained in this case (not displayed), 
that is, a ever drecreasing function in an expanding 2D universe. 
The results related
to the acceleration, in figure  (5)  show an strong dependence on fine tuning
of the values of parameters $\alpha$, $b$ and $\omega$ and more importantly
 it is 
evident that
the solely presence of the non-linear term does not guarantee   a 
final accelerated 
era for the 2D universe.

Next we investigate the model when the JT limit is not taken into account; 
that means, once one of the roots of equation (4) is chosen the  sign is 
mantained for all times (independently of the sign of the
trace $T$). The main point here is that in this 
case we get similar  transitions to the ones found in \cite{Deve}, a 
three-eras regime, and more interestingly  is that
 the non-linear term is effectively
promoting a final accelerated regime, as it is explained bellow.

\begin{figure}[h]
  \centering
  \mbox{
    \subfigure[]
    {\epsfig{figure=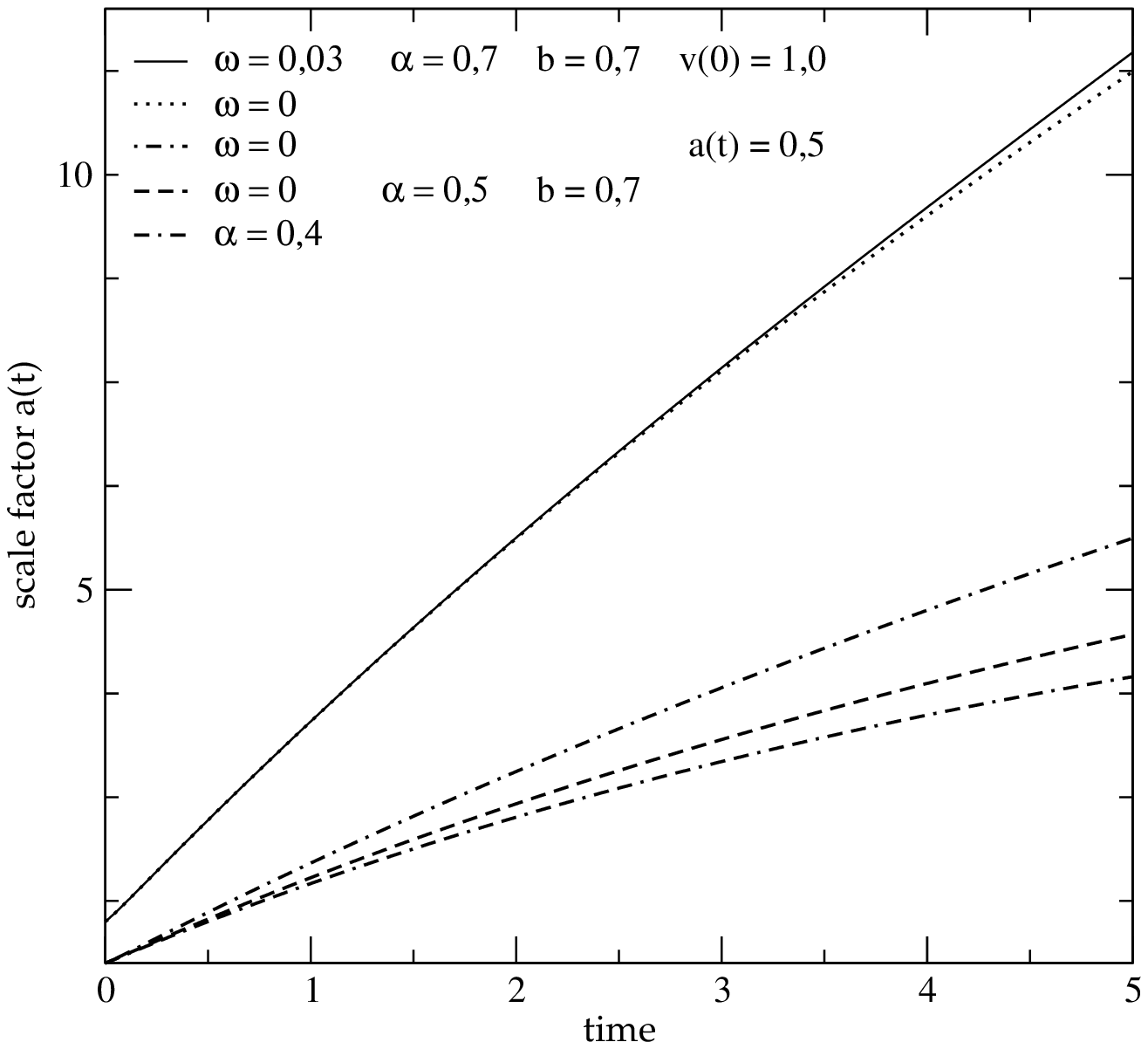,height=5.4cm,width=6.9cm}}
    \subfigure[]
    {\epsfig{figure=9.eps,height=5.4cm,width=6.9cm}}
    }\\
 \caption{Evolution in time for scale factor $a(t)$ and energy density $\rho $,
 for different values of
the parameters, without JT limit.}
\label{fig6}
 \mbox{
    \subfigure[]
    {\epsfig{figure=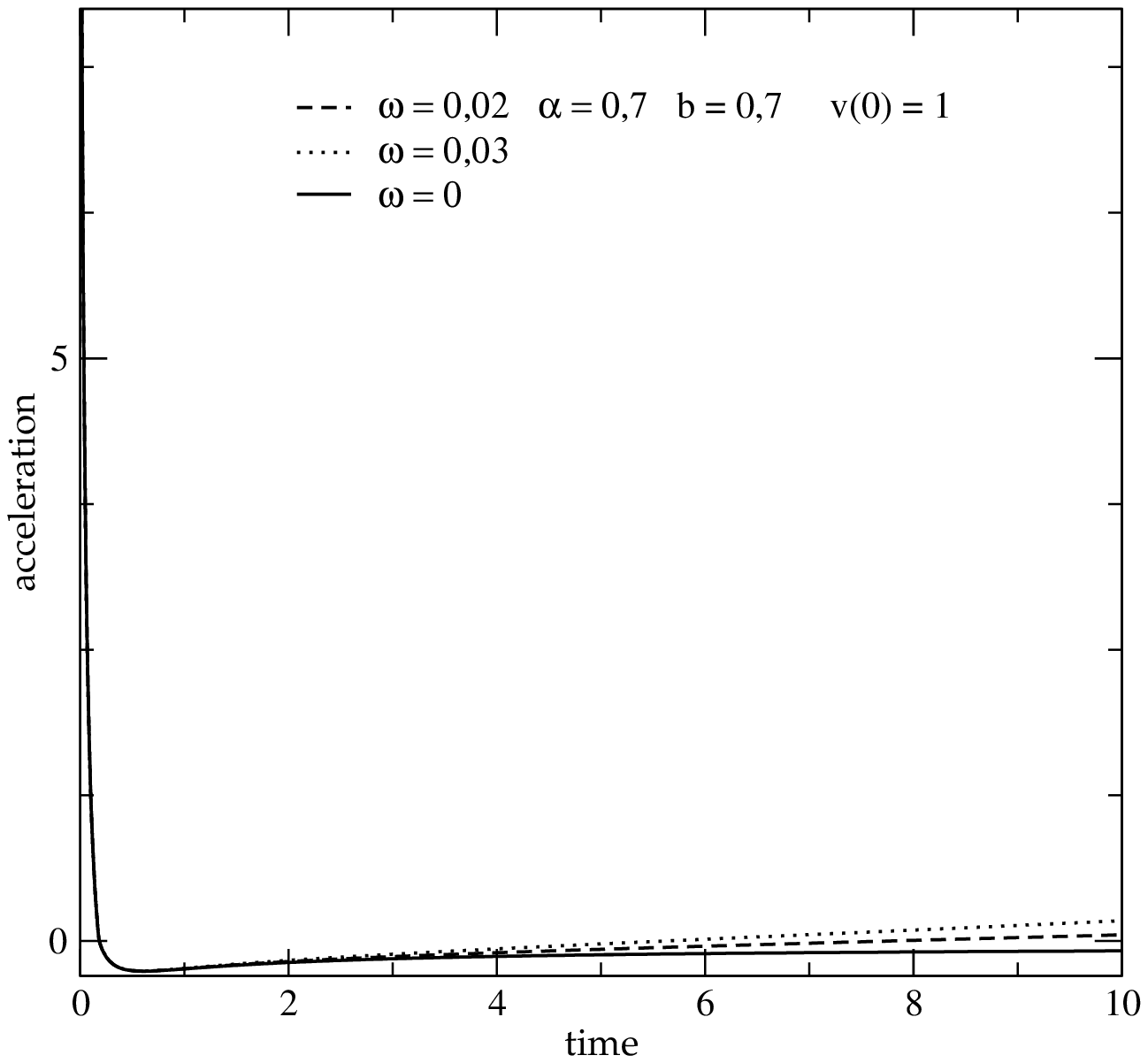,height=5.4cm,width=6.9cm}}
    \subfigure[]
    {\epsfig{figure=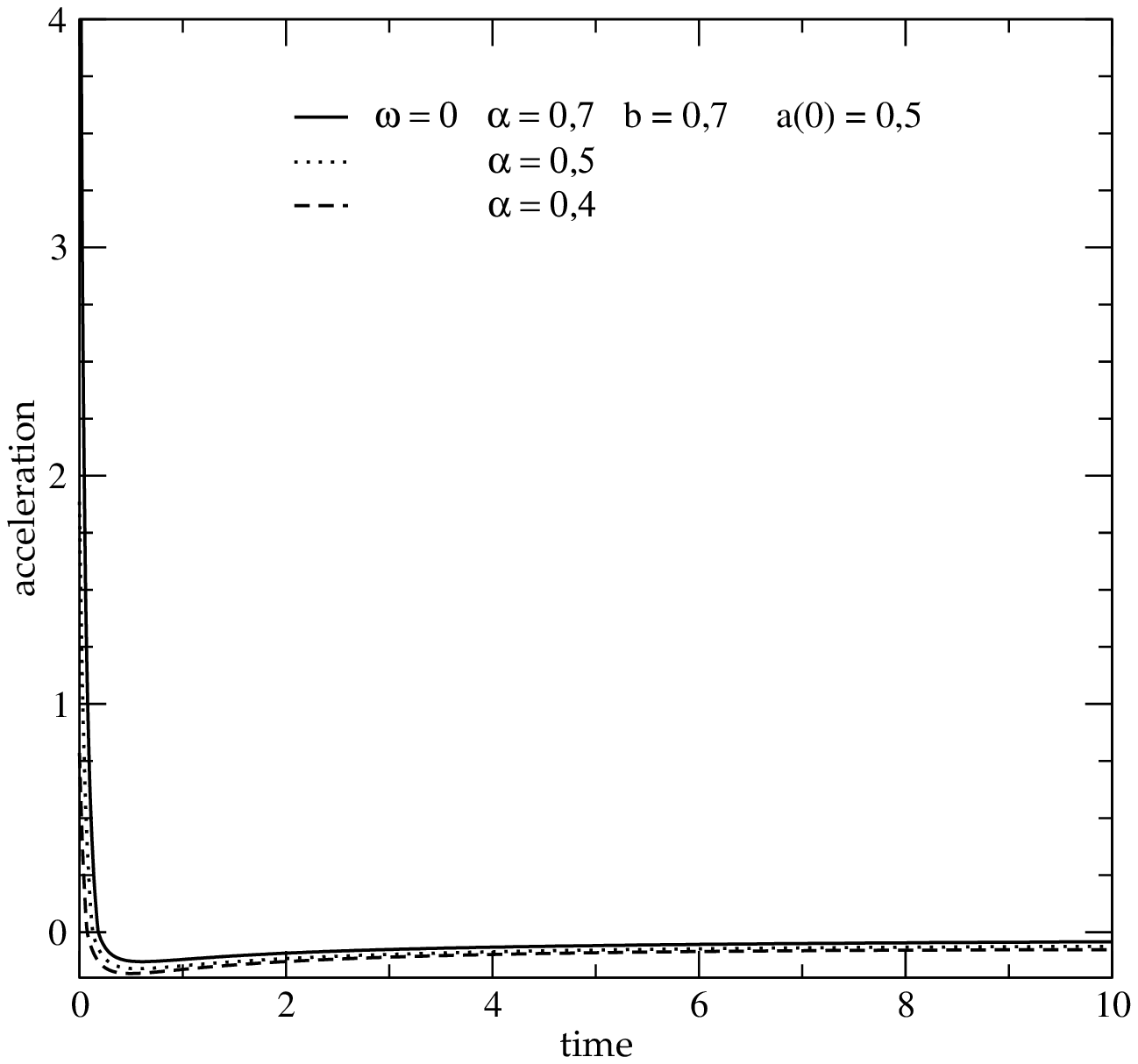,height=5.2cm,width=6.8cm}}
    }\\
 \caption{Evolution in time for acceleration $\ddot{a}(t)$,
 for different values of the parameters, without JT limit.}
\label{fig7}
\end{figure}

In figure (\ref{fig6})-(a) we display the behavior of the scale factor $a(t)$. 
The results
show  an ever expanding 2D universe, with increasing values of $\alpha$ and 
$b$ furnishing a faster expansion as in the previous case. Figure 
(\ref{fig6})-(b) 
shows (again) an expected behavior 
for the vdW energy density; increasing values of parameters $\alpha$ and
$b$ furnish a more drastic fall of the curve with the passing of time. Only
for later times the presence of the non-linear term in $R$ promotes a sensitive
increasing in the expansion rate of this 2D universe.

The behavior of the acceleration is the most interesting result, 
as we display in figure (\ref{fig7}). 
As the values of time increase the 
term in $1/R$ becomes more proeminent and boosts a transition between a 
decelerated era (dominated by matter, as far as
the vdW equation of state is becoming a barotropic one as the system evolves) 
and a 
accelerated period. This indicates that the non-linear term is playing a
role similar to  the cosmological constant (see \cite{Deve}) but 
 in this case the  responsible for that final 
accelerated era has a geometrical origin.  The effect happens 
in a more drastic way as parameter $\omega$  increases. Besides, increasing 
values of  parameters b and $\alpha$
promote an earlier transition to the final accelerated regime.
Although  the case with the JT limit is strongly suceptible to changes in the 
initial   
conditions the case without the JT limit
also shows situations where the association of the non-linear term with 
the dark energy is not possible. For instance,
for the conditons $ a(0)=1 \, v(0)=.9  $ and $ \rho (0)=.9$  a ciclic 2D universe 
emerges with a similarly ciclic behavior of
the acceleration $\ddot a(t)$. Finally we would like to articulate some comments on the results obtained in this work.
First,  we should mention that the scalar theory (in the form of the JT model) is a consistent theory of gravity only 
in 2D and hence the
 properties of the cosmological scenarios that emerge  can be exclusively 
associated   to 2D space-times. 
In the 4D case several experimental tests show that a scalar formulation ($R\sim T$) doesn't work even in considerable 
simple situations\cite{Cornish}. One  alternative is the Brans-Dicke formulation that 
includes an additional  field in
the form of the dilaton (there is a 2D counterpart to Brans-Dicke that is known as the CGHS model\cite{Brown}; the 
cosmological
results are of difficult interpretation in that case \cite{Deve}). Second, the substitution of the cosmological
constant with the non-linear term in $R$ is  interesting because it avoids the problem, that is  exclusive of 2D 
gravity, 
of modelling the dark energy
contribution  as a source with a negative energy term. Third, a consistent quantum analysis of
2D gravitational models exists\cite{Brown} and therefore  quantum effects can be included in a full quantum scenario
or in a semi-classical cosmological formulation\cite{Brown}; these subjects are under 
investigation by the authors.

\section{ Conclusions }
In this work we have investigated the possibility of a term of the form $1/R$ (as 
a correction to the Jackiw-Teitelboim equations) 
 be responsible for an final accelerated regime in an old 2D universe. This
term is  substituing the usual 
idea of modelling  the dark energy as a usual source 
(like an scalar field or a 
cosmogical constant) by giving it a geometrical origin.  The model is rich 
enough to include several regimes, furnishing in special cases a three period behavior: 
a first 
strongly accelerated period corresponding to a young 2D universe, a second 
period dominated by matter/radiation and  a final transition when the non-linear term 
starts to show its effects and promotes
the desired final accelerated regime. On the other hand the model is very sensible 
 to 
changes in initial conditions and the solutions include regimes where the 2D cosmogical 
interpretations are obscure.

\end{document}